\documentclass[sn-mathphys]{sn-jnl}

\jyear{2021}%

\theoremstyle{thmstyleone}%
\theoremstyle{thmstyletwo}%

\theoremstyle{thmstylethree}%

\usepackage{amsmath}
\usepackage{graphicx}
\usepackage{amssymb,bm}
\usepackage{amsmath,amsthm}
\usepackage{epstopdf}
\usepackage{comment}
\usepackage{hyperref}
\hypersetup{
    colorlinks=true,
    linkcolor=blue,
    filecolor=magenta,
    urlcolor=cyan,
    citecolor=magenta
}
\newcommand{\be}{\begin{equation}} 
\newcommand{\ee}{\end{equation}}

\newcommand{\Pc}{{\cal P}}

\begin{document}

\title[]{First-order thermodynamics of scalar-tensor gravity}


\author[1,2]{\fnm{Serena} \sur{Giardino}}\email{serena.giardino@aei.mpg.de}

\author*[3]{\fnm{Andrea} \sur{Giusti}}\email{agiusti@phys.ethz.ch}

\affil[1]{\orgname{Max Planck Institute for Gravitational Physics (Albert Einstein Institute)}, \orgaddress{\street{Callinstra{\ss}e 38}, \city{Hannover}, \postcode{30167}, \country{Germany}}}

\affil[2]{\orgdiv{Institute for Theoretical Physics}, \orgname{Heidelberg University}, \orgaddress{\street{Philosophenweg 16}, \city{Heidelberg}, \postcode{69120}, \country{Germany}}}

\affil[3]{\orgdiv{Institute for Theoretical Physics}, \orgname{ETH Zurich}, \orgaddress{\street{Wolfgang-Pauli-Strasse 27}, \city{Zurich}, \postcode{8093}, \country{Switzerland}}}


\abstract{The first-order thermodynamics of scalar-tensor theory is a novel approach that exploits the intriguing relationship between gravity and thermodynamics to better understand the space of gravity theories. It is based on using Eckart’s first-order irreversible thermodynamics on the effective imperfect fluid describing scalar-tensor gravity and characterises General Relativity as an equilibrium state, and scalar-tensor theories as non-equilibrium states, naturally describing the approach to equilibrium. Applications of this framework to cosmology, extensions to different classes of modified theories, and the formulation of two complementary descriptions based on the notions of temperature and chemical potential all contribute to a new and unifying picture of the landscape of gravity theories.}

\keywords{modified gravity, scalar-tensor gravity, non-equilibrium thermodynamics, cosmology}



\maketitle

\section{Introduction}
\label{sec:1}
\setcounter{equation}{0}
Einstein's General Relativity (GR) is the most successful theory of gravity to date, in excellent agreement with experimental tests, especially within the realm of our Solar System \cite{Will:2014kxa}. However, there are still considerable motivations to explore extensions and modifications of gravity beyond GR \cite{Faraoni:2010pgm}. 

First, Einstein gravity is a non-renormalisable theory involving incurable divergences at high energies, which makes it very challenging to integrate it into the framework of quantum field theory that works fruitfully for all other fundamental interactions. This is the reason why the enormous efforts in constructing a consistent theory of quantum gravity have not yet achieved definitive success and several candidates for such a theory exist, at different stages of development. Some extensions of GR that include higher-order curvature invariants in the action, such as $R^2$ gravity, are indeed renormalisable. This stimulates the exploration of gravity theories that could alleviate GR's problems in the UV regime.

Second, very compelling motivations to explore modifications or extensions of GR come from cosmology. The standard cosmological model, $\Lambda$CDM, is built on the pillars of GR and the cosmological principle (the assumption of homogeneity and isotropy of our universe). While successful in explaining most features of the observable universe, $\Lambda$CDM additionally relies on the presence of two unknown components, namely the dark energy driving the present accelerated expansion of the universe ($\Lambda$ stands for the cosmological constant, the simplest proposal for dark energy) and dark matter (CDM stands for cold dark matter). Since GR is not as observationally well-tested on cosmological scales as it is within the Solar System, there is room for accommodating deviations from it in those regimes. 
Especially the unresolved issue of dark energy provides uniquely fertile ground to explore modified theories: the cosmological constant is an extremely fine-tuned attempt to solve the problem, and it is natural to ask whether modifying GR could get rid of the need to introduce such a mysterious component.

In this context, one of the most thoroughly explored approaches is that of extending GR by adding new degrees of freedom to the two tensor ones, like a scalar or vector field \cite{Capozziello:2011et}. Scalar-tensor theories, first explored in \cite{Brans:1961sx} and generalized in \cite{Bergmann:1968ve,Nordtvedt:1968qs,Wagoner:1970vr}, are the simplest possible extension and their long-lasting popularity is also motivated by the fact that scalar fields are ubiquitous in cosmological models: one example is the inflaton field, which is posited to drive inflationary expansion in the early universe. Other more baroque extended gravity theories exist, for example those involving vector fields or those breaking some fundamental assumption of GR. 

Because of the plethora of different theories of gravity, it is useful to construct a unifying perspective on them by thinking of a vast landscape, where GR occupies a central place. Within this perspective, studying extended theories of gravity helps to understand GR as a special case in a much more general framework, which is a promising way to better understand and eventually overcome its limitations.

This is precisely the goal of the approach that we dub ``first-order thermodynamics of scalar-tensor gravity'', first developed in \cite{Faraoni:2021lfc} and extended in \cite{Giusti:2021sku,Faraoni:2021jri,Giardino:2022sdv,Faraoni:2022doe,Faraoni:2022gry,Giusti:2022tgq,Giardino:2023qlu}. 
This proposal is based on the observation that the contribution of the scalar field $\phi$ to the field equations of scalar-tensor gravity can be described as an effective 
dissipative fluid, through a simple rewriting of the equations that does not entail extra assumptions \cite{Faraoni:2018qdr,Pimentel:1989bm}. The novelty of this approach comes in when we apply a non-equilibrium thermodynamical description to this fluid. We choose the one developed by Eckart \cite{Eckart:1940te}, entailing constitutive relations that are first-order in the dissipative variables and, despite problems related to causality that stem from this simplistic assumption, is still one of the most widely used models of dissipative thermodynamics. Giving a thermodynamical interpretation to the effective $\phi$-fluid leads to the identification of its temperature, a sort of ``temperature of scalar-tensor gravity'', which is nothing but a temperature relative to GR, in addition to its bulk and shear viscosity coefficients.
This temperature is the order parameter ruling the approach to equilibrium and it is positive definite for theories containing a scalar degree of freedom in addition to the two tensor ones of GR: Einstein's theory corresponds to the zero-temperature state in this ``thermodynamics of gravitational theories''. Dissipation corresponds to the relaxation of
the effective fluid toward the GR state of equilibrium. This approach fits into the wider context of trying to gain physical intuition through an effective fluid description for theories involving complicated derivative self-interaction terms in their Lagrangians (see, for example, \cite{Pujolas:2011he,Miranda:2022wkz}).

Our approach is inspired by and echoes the ideas of \cite{Jacobson:1995ab,Eling:2006aw}, but follows a starkly different path. These previous works derived both Einstein's equations and the field equations of $f({\cal R})$ gravity (a subclass of scalar-tensor theories \cite{Sotiriou:2008rp}) as equations of state from purely thermodynamical considerations, also leading to the identification of GR with an equilibrium state of gravity and modified gravity with a non-equilibrium one. These results made the interesting relationship between gravity and thermodynamics, originally explored in the context of black holes, even more intriguing. However, they left open the crucial questions of a precise description of the approach to equilibrium and the order parameter governing it, which our approach addresses.

The present work is structured as follows: Sec.~\ref{generalities} provides the general framework of first-order thermodynamics, in Sec.~\ref{otherequi} we study equilibrium states different than GR with the goal of assessing their relevance and stability, Sec.~\ref{extensions} addresses some applications and extensions of the formalism, such as an alternative picture in Sec.~\ref{chemical}, the application to cosmology in Sec.~\ref{cosmology} and to Horndeski theories in Sec.~\ref{horndeski}.

We adopt the notation of Ref.~\cite{Waldbook}: the metric signature is $(-,+,+,+)$ and we use units in which the speed of light $c$ and Newton's constant $G$ are unity.

\section{Effective fluid formalism and Eckart's thermodynamics}
\label{generalities}
The effective fluid description relies on a single assumption, namely that the gradient of the scalar field $\nabla^a\phi$ is timelike and future-oriented \cite{Giusti:2022tgq}, so that a 4-velocity for the fluid can be meaningfully defined as
\be
\label{velocity}
u^a=\frac{\nabla^a\phi}{\sqrt{-\nabla^e\phi\nabla_e\phi}},
\ee
where $u^a u_a=-1$. This allows the $3+1$ splitting of spacetime into the time direction given by $u^a$ and the 3-dimensional space of observers comoving with the fluid and the definition of the induced metric
\be
h_{ab}=g_{ab}+u_a u_b,
\ee
so that ${h_a}^d$ is the projection operator on this 3-space that allows us to define kinematic quantities such as the expansion tensor $\Theta_{ab}$ and the shear tensor $\sigma_{ab}$ (see \cite{Faraoni:2021jri} for details).

If the fluid velocity is past-oriented, on the other hand, the kinematic fluid quantities remain unchanged, but certain thermodynamical variables such as the heat flux change sign, leading to a negative temperature and a positive shear viscosity, at variance with previous works. Given that a negative temperature is meaningless, the formalism presented here is only valid for future-oriented fluid velocity \cite{Giusti:2022tgq}.

Let us now see how the effective fluid formalism can be employed, starting from the scalar-tensor action (in the Jordan frame) 
\be 
S_\text{ST} = \frac{1}{16\pi} \int 
d^4x \sqrt{-g} \left[ \phi {\cal R} -\frac{\omega(\phi )}{\phi} \, 
\nabla^c\phi \nabla_c\phi -V(\phi) \right] +S^\text{(m)} \,, 
\label{STaction} 
\ee 
where ${\cal R}$ is the Ricci scalar, the Brans-Dicke 
scalar $\phi>0$ is approximately the inverse of the effective gravitational 
coupling $G_\mathrm{eff}$, $\omega(\phi)$ is the ``Brans-Dicke coupling'', 
$V(\phi)$ is the scalar field potential, and $S^\text{(m)}=\int d^4x \sqrt{-g} \, 
{\cal L}^\text{(m)} $ is the matter action.

The field equations are \cite{Brans:1961sx,Wagoner:1970vr,Bergmann:1968ve} 
\begin{eqnarray} 
G_{ab} &\equiv & {\cal R}_{ab} - \frac{1}{2}\, g_{ab} {\cal R} = 
\frac{8\pi}{\phi} \, T_{ab}^\text{(m)}
+ \frac{\omega}{\phi^2} \left( \nabla_a \phi \nabla_b \phi -\frac{1}{2} \, 
  g_{ab} \nabla_c \phi \nabla^c \phi \right) \nonumber\\
&&\nonumber\\
&\,& +\frac{1}{\phi} \left( \nabla_a \nabla_b \phi
- g_{ab} \Box \phi \right) -\frac{V}{2\phi}\, g_{ab} \,  \label{BDfe1} \\
&&\nonumber\\
\label{BDfe2}
\Box \phi &=& \frac{1}{2\omega+3} \left( \frac{8\pi T^\text{(m)} 
}{\phi} + \phi \,  V_{,\phi}
-2V -\omega_{,\phi} \, \nabla^c \phi \nabla_c \phi \right) \,, 
\end{eqnarray} 
where ${\cal R}_{ab}$ is the Ricci tensor, $ T^\text{(m)} 
\equiv g^{ab}T_{ab}^\text{(m)} $ is the trace of the matter stress-energy 
tensor $T_{ab}^\text{(m)} $, and $\omega_{,\phi} \equiv d\omega/d\phi $, 
$ V_{,\phi} \equiv dV/d\phi$. 

The stress-energy tensor of the effective fluid represented by the scalar contributions can be read off the right-hand side of Eq.~(\ref{BDfe1})
\begin{eqnarray} 
\nonumber
8\pi T_{ab}^{(\phi)} &=& \frac{\omega}{\phi^2} \left( \nabla_a \phi 
\nabla_b \phi -
 \frac{1}{2} \, g_{ab} \nabla^c \phi \nabla_c \phi \right)  + 
 \frac{1}{\phi} \left( \nabla_a \nabla_b \phi -g_{ab} \square \phi \right)
- \frac{V}{2 \phi} \, g_{ab}. \\
&& \label{BDemt} 
\end{eqnarray} 
$T_{ab}^{(\phi)}$ has the form of an imperfect fluid energy-momentum  
tensor \cite{Pimentel:1989bm,Faraoni:2018qdr}, 
\be 
T_{ab} = \rho \, u_a u_b + q_a u_b + q_b u_a + \Pi_{ab} 
\,,\label{imperfectTab} 
\ee 
where the comoving effective energy density, heat flux density, stress tensor, 
isotropic pressure, and anisotropic stresses (the trace-free part 
$\pi_{ab}$ of the stress tensor $\Pi_{ab}$) are, respectively,  
\begin{align} 
\rho &= T_{ab}    u^a u^b \, , \label{rhophi}\\ 
q_a &= -T_{cd} \, u^c 
{h_a}^d  \, ,
  \label{qphi}\\
 \Pi_{ab} &= Ph_{ab} + \pi_{ab} = T_{cd} \, {h_a}^c \, {h_b}^d \, , 
\label{Piphi}\\
    P &= \frac{1}{3}\, g^{ab}\Pi_{ab} =\frac{1}{3} \, h^{ab} T_{ab}  \, ,
\label{Pphi}\\
    \pi_{ab} &= \Pi_{ab} - Ph_{ab} \, \label{piphi}.
\end{align}

Note that the decomposition in \eqref{imperfectTab} applies to any symmetric second-order tensor, although of course the dissipative quantities would vanish if the effective stress-energy tensor for the theory at hand takes the form of a perfect fluid (see Section \ref{chemical} for more details). The special feature of scalar-tensor gravity in first-order thermodynamics is not that the decomposition above can be performed, but rather that the constitutive relations of Eckart's thermodynamics hold \cite{Faraoni:2023hwu}.

These constitutive relations are the simplest (linear) assumptions to satisfy $\nabla_{\alpha}S^{\alpha}\geq0$, the covariant second law of thermodynamics \cite{Eckart:1940te,Maartens:1996vi}. They relate the viscous pressure $P_\text{vis}^{(\phi)}$ to the expansion scalar $\Theta$ through the bulk viscosity coefficient $\zeta$, the heat current density 
$q_a^{(\phi)}$ to the temperature ${\cal T}$ through the thermal conductivity ${\cal K}$, and the anisotropic stresses $\pi_{ab}^{(\phi)}$ through the shear viscosity coefficient $\eta$ to the shear tensor $\sigma_{ab}$. For the $\phi$-fluid, they read:
\begin{align}
\label{EckartPvis}
P^{(\phi)}_\text{vis} &= -\zeta \, \Theta 
\\
\label{Eckartq}
q_a^{(\phi)} &= -{\cal K} \left( h_{ab} \nabla^b {\cal T} + {\cal T} \dot{u}_a \right) \\
\label{Eckartpi}
\pi_{ab}^{(\phi)} &= - 2\eta \, \sigma_{ab} \,. 
\end{align} 
They are relativistic generalisations of Stokes' law, Fourier's law and Newton's law of viscosity, respectively.

A simple comparison between the expressions 
of the 4-acceleration, $\dot{u}_a\equiv u^b\nabla_b u_a$ and the heat flux density $q^{(\phi)}_a$ 
for scalar-tensor theories described by the action \eqref{STaction} leads to the identification \cite{Faraoni:2018qdr,Faraoni:2021lfc,Faraoni:2021jri}
\be 
q_a^{(\phi)} = -\frac{ \sqrt{-\nabla^c \phi \nabla_c \phi}}{ 8 \pi \phi} 
\,  \dot{u}_a
\label{q-a} 
\ee 
and comparison with Eckart's generalized Fourier law \eqref{Eckartq} yields an expression for the product of thermal conductivity and temperature,
\be
{\cal K} {\cal T} = \frac{ \sqrt{-\nabla^c \phi \nabla_c \phi}}{ 8 \pi 
\phi}.  \label{temperature}
\ee
Additionally, one finds $h_{ab}\nabla^b {\cal T}=0$, which means that a heat flux arises from accelerated matter even in absence of a temperature gradient, as first identified precisely by Eckart \cite{Eckart:1940te}.
What is most striking is that ${\cal KT}$ is positive definite and therefore meaningful, which was not to be expected \textit{a priori} in a formal identification of quantities. Moreover, ${\cal KT}$ vanishes when $\phi=\rm const.$, namely in the GR limit, where there is no $\phi$-fluid.

GR can therefore be identified with the ${\cal KT}=0$, equilibrium state in this thermodynamics of gravitational theories, whereas in general scalar-tensor theories have ${\cal KT}>0$. This is the sense in which first-order thermodynamics provides a picture of the landscape of gravity theories, namely by putting GR as an equilibrium state at the centre of a space of theories populated by scalar-tensor theories, representing non-equilibrium states. 
This echoes the approach in \cite{Jacobson:1995ab,Eling:2006aw} where a non-equilibrium thermodynamical setting was required to deal with a modified theory of gravity. 

A simple physical interpretation of the quantities ${\cal K}$ and ${\cal T}$ arises if we isolate the temperature from \eqref{temperature} and insert it into $h_{ab}\nabla^b {\cal T}=0$, finding the simple solution ${\cal K}=C\sqrt{-\nabla^c \phi \nabla_c \phi}$, with $C$ a positive constant that can be set to $C=1/8\pi$, yielding ${\cal T}=1/\phi=G_{\rm eff}$. This sheds light on the GR limit, which recovers the ``perfect insulator'' limit of the effective fluid: ${{\cal T}}$ reduces to the Newton constant $G_{\rm N}$, while ${\cal K}$ vanishes.

The structure of the imperfect fluid \eqref{imperfectTab} and of the field equation \eqref{BDfe2} makes the explicit derivation of the bulk viscosity from the thermodynamic analogy (feasible but) nontrivial.
For the sake of simplicity we shall set the bulk viscosity to zero as in the original proposal \cite{Faraoni:2021lfc,Faraoni:2021jri}.\footnote{For a more precise analysis on this matter we refer the reader to \cite{Miranda:2022wkz}.}
Nonetheless, one can still easily infer the shear viscosity coefficient in a similar way as above, obtaining \cite{Faraoni:2021lfc,Faraoni:2021jri}
\be
\label{eta}
\eta=-\frac{ \sqrt{-\nabla^c \phi \nabla_c \phi}}{ 16 \pi \phi}
\ee
or $\eta=-\frac{{\cal KT}}{2}$. The viscosity can be negative as the $\phi$-fluid is clearly not isolated, given the explicit coupling to gravity in the action \eqref{STaction}, which involves a mixing of scalar and tensor degrees of freedom.

\subsection{Approach to thermal equilibrium}
\label{approachsection}
It is natural to ask how equilibrium might be approached starting from a non-equilibrium state, as the understanding of this dissipative process is crucial to establish the picture we are trying to construct.
An effective heat equation for the $\phi$-fluid can be found by differentiating \eqref{temperature}. Although this might seem redundant, the resulting equation provides the behaviour of ${\cal KT}$ with time and allows us to understand the circumstances where the dissipation to equilibrium takes place.
Differentiating $\frac{d({\cal KT})}{d\tau}\equiv u^c\nabla_c({\cal KT})$, we obtain \cite{Faraoni:2021lfc,Faraoni:2021jri}
\be
\label{approach}
\frac{d({\cal KT})}{d\tau}=8\pi({\cal KT})^2-\Theta ({\cal KT})+\frac{\Box\phi}{8\pi\phi}.
\ee
A general interpretation of this equation is challenging since $\Box\phi$ does not have definite sign and the dependence of $\Theta$ on $\phi$ and its derivatives is not straightforward, but one can gain some physical intuition by considering the vacuum case, with $\omega=\rm const.$ and $V(\phi)=0$, so that $\Box\phi=0$. 

On the one hand, if $\Theta<0$, ${\cal KT}$ grows out of control in a finite time and diverges away from equilibrium. This behaviour is relevant around spacetime singularities, where $\Theta<0$ because the worldlines of the field $\phi$ converge: in our formalism, this means that the deviations of scalar-tensor gravity from GR will be extreme. 

On the other hand, if $\Theta>0$, depending on which term dominates in \eqref{approach}, ${\cal KT}$ could either asymptote to zero and approach the equilibrium state, or not. Thus, the approach to equilibrium is not granted and cases where it does not happen will be considered in the following.

\section{Analysis of equilibrium states other than GR}
\label{otherequi}
We can gain more insight into the effective heat equation above by studying its fixed points, \textit{i.e.} when $\frac{d({\cal KT})}{d\tau}=0$. Of course, they correspond to situations where either ${\cal KT}=0$ or ${\cal KT}=\rm const.$ and constitute equilibrium states in the thermodynamics of gravitational theories. Our goal is thus to analyse these other possible equilibrium states that might challenge the uniqueness of the GR equilibrium state and its special role in this landscape of gravity theories. In turn, this will provide additional tests of our formalism and its physical interpretation. The thermodynamical quantities of the effective fluid  can of course be found starting from a different scalar-tensor action than \eqref{STaction}, but one can also specify the expressions such as ${\cal KT}$  to exact solutions of the theory at hand. The formalism is therefore flexible enough that we can study both entire classes of theories and specific solutions within it, which is what we show in the following.

\subsection{Zero-temperature states: non-dynamical scalars}
In \cite{Faraoni:2022doe}, we studied theories of gravity other than GR that have ${\cal KT}=0$, with the goal of further exploring the theory landscape through the lens of the formalism. The theories considered are not always physically viable, but they allow to clarify the existence of other equilibrium states and test the regime of validity of first-order thermodynamics. The theories with non-dynamical scalar fields that we considered are Brans-Dicke gravity \cite{Brans:1961sx} in the limit $\omega=-3/2$, $f(R)$ gravity in the Palatini formulation (equivalent to $\omega=-3/2$ Brans-Dicke with a specific potential \cite{Sotiriou:2008rp}), and cuscuton models \cite{Afshordi:2006ad} (which can be recast as special cases of Horndeski theory, discussed in the following). The first two have ill-defined or zero ${\cal KT}$, while the cuscuton case provides the opportunity for a richer analysis.

Additionally, by extending the formalism to the case of Nordstr\"om gravity \cite{nordstroem} (a purely scalar, unviable theory of gravity that only has historical importance as a stepping stone towards GR), we discovered that a theory with less degrees of freedom than GR has negative temperature in the formalism (and bulk and shear viscosity coefficients of opposite signs with respect to the standard case), further validating the hypothesis that a positive temperature is tied to the existence of an additional scalar degree of freedom to the two tensorial ones. Of course, a negative temperature is devoid of meaning and makes the formalism as pathological as the theory itself. 

Let us briefly review the case of the cuscuton model \cite{Afshordi:2006ad}, which corresponds to a special case of the Lorentz-violating Ho\v{r}ava-Lifshitz gravity \cite{Afshordi:2009tt} and has intriguing cosmological implications \cite{Afshordi:2007yx,Quintin:2019orx}. The cuscuton field (denoted as $\phi$ in the following) is a scalar that does not propagate additional degrees of freedom with respect to GR and whose equation of motion is shown to reduce to a constraint.
The cuscuton action is
\be
S_{\rm C}= \int d^4 x \sqrt{-g} \,\left( \frac{{\cal R}}{16\pi} +\Pc \right) 
+S^{(\rm m)} \,,
\ee
where the Lagrangian density is equivalent to the pressure that we denote with $\Pc$ in this section, is \cite{Quintin:2019orx}
\be
\Pc (\phi, X) = \pm \mu^2 \sqrt{2X} -V(\phi) \,,
\ee
$X \equiv -\frac{1}{2}\nabla^c \phi \nabla_c \phi$, $\mu$ is a mass scale and in the following $f_{X}\equiv \frac{\partial f}{\partial X}$.
The effective stress-energy tensor appearing in the Einstein equations, namely
\begin{eqnarray}
T_{ab}^{(\phi)} &=& \Pc g_{ab} + \Pc _{X} \nabla_a \phi \nabla_b \phi =\left[ \pm \mu^2 \sqrt{2X} -V(\phi) \right] g_{ab} \pm 
\mu^2 \, \frac{\nabla_a \phi \nabla_b \phi}{\sqrt{2X}}, 
\end{eqnarray} 
takes the perfect fluid form $T_{ab}=\left( \Pc+\rho \right) u_a u_b +\Pc g_{ab}$ (in contrast to \eqref{imperfectTab}), so that no imperfect fluid description can arise and, accordingly, ${\cal KT}=0$.

These results confirm the intuition that theories of gravity with dynamical scalar fields have ${\cal KT}>0$, while those with non-dynamical fields have either zero or completely arbitrary ${\cal KT}$. The theories with ${\cal KT}=0$ that we analysed are peculiar and/or pathological and cannot compete with the central role of GR as an equilibrium state at ${\cal KT}=0$.

\subsection{Constant temperature states: stealth solutions}
The so-called stealth solutions commonly arise in scalar-tensor theories: they have the same geometry of GR solutions but a nontrivial scalar field profile that does not contribute to the effective stress-energy tensor. Currently, the motivation to study them comes from the fact that such solutions can violate some assumptions of the black hole no-hair theorem, which would make it possible in principle to observationally distinguish GR from scalar-tensor theories through the detection of gravitational waves from black hole mergers.

In \cite{Faraoni:2021jri}, it was noticed that a stealth solution of scalar-tensor gravity was characterized by ${\cal KT}=\rm const.$, corresponding to a fixed point of \eqref{approach}. This is interesting because it would mean that stealth solutions are examples of states that never approach the GR equilibrium. In \cite{Faraoni:2022jyd}, this solution was studied in more detail and found to be metastable and in \cite{Giardino:2023qlu} we undertook a more extensive analysis of stealth solutions in our formalism. 
In order to assess the relevance of the equilibrium states corresponding to ${\cal KT}=\rm const.$, it is essential to check their stability with respect to some criterion. Of course, in different systems, distinct types of stability can arise (thermal, dynamical, {\em etc.}) which are not necessarily expected to coincide: in the following, we make use of different types of stability criteria. If the equilibrium states we study are not stable, it means they cannot compete with the zero-temperature state constituted by GR, further strengthening its special role. Of course, it is not practically feasible to study all possible equilibrium states and, for the time being, first-order thermodynamics is not in the position to turn the statement that GR is the only possible equilibrium state into a formal theorem. Therefore, the nature of this statement is an inductive result based on the most relevant theories that we studied.

In \cite{Giardino:2023qlu} we considered stealth solutions where Minkowski geometry results from a tuned balance between matter and the scalar field or, in a vacuum configuration, between different terms in the scalar contribution to the stress-energy tensor. These solutions are often degenerate de Sitter spaces with non-constant 
scalar fields, which are cannot be realized in GR and constitute a signature of modified gravity.

In order to study the stability of these solutions, we develop a criterion based solely on the thermodynamical formalism. In general, this criterion should be seen as complementary to others, such as the gauge-invariant one developed for cosmological perturbations \cite{Bardeen:1980kt, Ellis:1989jt, Ellis:1990gi, 
Hwang:1990am}.

The criterion is obtained from rewriting \eqref{approach} as
\be
\label{effkg}
\Box \phi-m_{\rm eff}^2 \phi=0,
\ee
where
\be
m_{\rm eff}^2 \equiv 8\pi \left[ \frac{d\left( {\cal KT}\right)}{d\tau} - 
8\pi 
\left( {\cal KT}\right)^2 + \Theta \, {\cal KT} \right] \,. 
\label{criterion}
\ee 
One can easily see that if the square of the ``effective mass'' is $ m_{\rm eff}^2<0$, this tachyonic behaviour signals an instability and renders the solution at hand problematic, while there is stability if 
$ m_{\rm eff}^2 \geq 0$. Of course, these considerations fully rely on first-order thermodynamics only and are therefore meaningful only within the formalism. Since ${\cal KT}$ is a scalar, this stability criterion is covariant and gauge-invariant.

This criterion is not particularly 
useful for entire classes of scalar-tensor theories because the quantities appearing in (\ref{criterion}) are not necessarily known, but it is suitable for the study of specific solutions (or classes of solutions) of the field equations in these theories, which is why we use it for stealth solutions.

Stealth solutions commonly encountered in the literature in the context of 
the scalar-tensor theory described by (\ref{STaction}) are usually  
of two kinds:

\begin{enumerate} 

\item $g_{ab}=\eta_{ab} $ and $\phi=\phi_0 \, e^{\alpha \, t} $;

\item $g_{ab}=\eta_{ab} $ and $\phi=\phi_0 \, \(\lvert t\rvert\)^{\beta} $,

\end{enumerate}
\noindent where $\eta_{ab}$ is the Minkowski metric, $\phi_0, \alpha, \beta$ are constants, and $\phi_0>0$. 
Keeping in mind that the scalar field gradient needs to be future-oriented for the 4-velocity of the effective fluid (and consequentially, for the whole formalism) to be meaningful \cite{Giusti:2022tgq}, we restrict to cases satisfying the conditions
\begin{enumerate} 
\item $\alpha < 0$;
\item $\beta<0 $~ if $t>0$ or ~$\beta>0$~ if $ t<0$.
\end{enumerate}
In the first case
\be
{\cal KT} = \frac{ \sqrt{-\nabla^c \phi \nabla_c\phi}}{8\pi \phi} = 
\frac{\(\lvert \alpha\rvert\)}{8\pi} =\mbox{\rm const.}>0 \,, 
\label{stealth-exp}
\ee
so this solution can never approach the GR equilibrium state. Applying the criterion \eqref{criterion}, we find a constant and negative effective mass, meaning that this stealth solution is unstable.

In the second case, we restrict to $\beta=1$ and $\beta=2$, which are the most relevant cases in the literature. 
We find
\be
{\cal KT} = \frac{ 
\beta}{8\pi \(\lvert t\rvert\)} \to 
+\infty \quad 
\mbox{as} \; t\to 0^{-} \, ,
\ee
for the effective temperature, while the effective mass reads
\be
m_\mathrm{eff}^2 =  \frac{\Box\phi}{\phi} = - \frac{\beta (\beta - 1)}{t^2}  \, .
\ee
The case of $\beta=1$ corresponds to marginal stability since $m_{\rm eff}^2 = 0$, while $\beta=2$ entails the instability corresponding to $m_\mathrm{eff}^2 = - 2/t^2 <0$. In both cases, as $t\to 0^{-}$, we approach a singularity of the theory where $G_\mathrm{eff}$ diverges just as ${\cal KT}$ does: gravity becomes infinitely strong and 
deviates from GR drastically, which is the expected behaviour for singularities in first-order thermodynamics, as explained in Section \ref{approachsection}.

We applied these general considerations to some specific stealth solutions \cite{Giardino:2023qlu}, including some specific cases of exact solutions of FLRW cosmology in scalar-tensor theories and degenerate cases of de Sitter solutions that reproduce Minkowski space with a nontrivial scalar field. The result is that no stable states of equilibrium have been found, except for solutions that are unstable according to 
various criteria and thus irrelevant. This result strengthens the special role of GR as an equilibrium state in the landscape of gravity theories, seen through the lens of first-order thermodynamics.

\section{Further extensions of the formalism}
\label{extensions}
\subsection{Alternative formulation with chemical potential}
\label{chemical}
The entire thermodynamical formalism so far has been developed starting from the Jordan frame action \eqref{STaction}, where there is an explicit coupling between the Ricci scalar and the $\phi$ field. However, scalar-tensor theories can also be studied in the (conformally related) Einstein frame, where the scalar couples minimally to gravity but nonminimally to matter. Switching to the Jordan frame amounts to performing the conformal transformation $g_{ab} \to \tilde{g}_{ab} \equiv \phi \, g_{ab}$, together with the field redefinition $\phi \to \tilde{\phi}$, where
$d\tilde{\phi} = \sqrt{ \dfrac{\(\lvert 2\omega+3\rvert\)}{16\pi}} \, \dfrac{d\phi}{\phi} \,$
and quantities with tilde refer to variables in the Einstein frame.
The action for GR with a minimally coupled scalar reads
\be
S_{\rm min}=\int d^4 x \, \sqrt{-g} \left[ \frac{{\cal R}}{16\pi} +
{\cal L}\left( \phi, X \right) 
\right] +  S^\mathrm{(m)} \,.
\ee
The thermodynamical formalism based on the notion of temperature could not be applied to theories in the Einstein frame, since the minimally coupled scalar gives rise to an effective fluid that is perfect. Since all imperfect fluid quantities vanish, the analogy built in the previous sections becomes trivial: ${\cal KT}$ is always zero and no approach to equilibrium (or departure from it) can be analysed.

However, in \cite{Faraoni:2022gry}, it was realised that an alternative but equivalent picture of first-order thermodynamics, based instead on the notion of chemical potential, can address this problem. Although 
${\cal KT}$ vanishes for the fluid describing the minimally coupled scalar, the chemical potential, defined as
$ \tilde{\mu} = \sqrt{ 2\tilde{X}}$ does not, and the dissipation to equilibrium can be characterised as $\tilde{\phi}\to~\rm const.$ and $\tilde{\mu}\to 0$.

This approach is reminiscent of the influential one in \cite{Pujolas:2011he}, that exploited the analogy between an imperfect fluid and a special class of scalar-tensor theories to help gain physical insight into such theories and their interesting cosmological applications.  

If ${\cal L}={\cal L}(X)$, the theory is invariant under the shift 
symmetry $\phi \to \phi +C$, where $C$ is a constant. This means there is a conserved Noether current $N^a = {\cal L}_X \nabla^a \phi = n u^a$, which satisfies 
$\nabla_a N^a=0$, where $N^a$ is the analogue of the particle number current density and $n$ is the particle number density in the comoving frame and corresponds to the Noether charge 
\be 
n = -N^0 = -u^c N_c = -\frac{\nabla^a \phi}{\sqrt{2X}} \, {\cal L}_X \nabla_a \phi = 
\sqrt{2X} \, {\cal L}_X \,. 
\ee 
However, in Eckart's thermodynamics, which corresponds to choosing the reference frame comoving with the fluid, no flux of ``$\phi$-particles'' is visible because this frame follows the effective fluid motion (for a discussion of frame choices and their physical interpretation, see \cite{Miranda:2022uyk}).

\subsection{Application to cosmology}
\label{cosmology}
Cosmology is the natural arena to work with scalar-tensor theories, not only because on cosmological scales there is still room to accommodate deviations from GR, but also since these theories were formulated in the first place to incorporate Mach's principle and to allow for a variation of the gravitational coupling, meaning that the distribution of matter on cosmological scales could have effects on local gravity \cite{Brans:1961sx}.

In \cite{Giardino:2022sdv}, we applied first-order thermodynamics to a cosmological setting by restricting to FLRW spacetime, whose symmetries (homogeneity and isotropy) allow for some simplifications in the thermodynamical quantities. We subsequently tested our physical intuition of the formalism on some exact solutions of scalar-tensor cosmology. The main result is that the GR equilibrium
state of zero temperature is almost always approached at late times throughout the universe's expansion, while the behaviour expected for singularities is confirmed for solutions endowed with a singularity at early times.

Because of homogeneity 
and isotropy, in FLRW $\phi= \phi(t)$, the heat flux density 
$q_a^{(\phi)} = 0$ and the anisotropic stresses $\pi_{ab}^{(\phi)} = 0$.
Therefore, shear viscosity vanishes, but the isotropic bulk viscosity that was previously neglected can be considered, and the only two non-vanishing contributions to \eqref{imperfectTab} are the isotropic pressure and energy density. The viscous contribution to the pressure can be isolated from the non-viscous terms by noticing that, in a FLRW spacetime, the expansion scalar is $\theta=3H$ and thus
\be
P^{(\phi)}_\mathrm{vis} = -3 \zeta H,
\label{Pvisc}
\ee
according to Eckart's constitutive relation~(\ref{EckartPvis}).
The bulk viscosity coefficient is identified as
\be
\zeta = \frac{\dot{\phi}}{24\pi \phi} \,\label{zeta},
\ee
while
\be
{\cal K T}= \frac{\lvert\dot{\phi}\rvert}{8\pi \phi}.
\ee
The bulk viscosity coefficient $\zeta={\cal KT}/3$ then scales linearly with the temperature and both vanish in the GR 
equilibrium state. 

These general formulas have been applied in \cite{Giardino:2022sdv} to the specific universe models described by exact solutions of scalar-tensor 
cosmology that generally have a power-law behaviour in time both for the scale factor and the scalar field \cite{mybook1}, such as the Brans-Dicke dust solution \cite{Brans:1961sx} and the Nariai family of solutions \cite{Nariai} (see also \cite{Faraoni:2022fxo}). We found that these solutions all approach the GR equilibrium state at ${\cal KT}=0$ and $\zeta=0$ at late times, in the limit $t\to +\infty$. A Big Rip solution \cite{Faraoni:2003jh} exhibiting a singular behaviour at late times with the scale factor diverging at a finite time proved puzzling because of the competition between the singularity behaviour ${\cal KT}\to + \infty$ and the behaviour at late times. In the end, we found that gravity still departs from equilibrium in the Big Rip solution.

\subsection{Extension to Horndeski theories}
\label{horndeski}
An application of the thermodynamical formalism that yielded compelling results was that to Horndeski gravity \cite{Giusti:2021sku}, the most general class of scalar-tensor theories exhibiting second-order equations of motion and thus avoiding Ostrogradsky instabilities. Horndeski theories represent a cornerstone of the literature on modified gravity which has seen a flurry of activity in recent years \cite{Kobayashi:2019hrl}. Such theories were employed in models of dynamical dark energy and as late-time modifications of GR, but have been severely constrained by the multi-messenger gravitational wave event GW170817 \cite{Ezquiaga:2018btd} that showed that gravitational waves propagate at the speed of light up to remarkable precision. 

The thermodynamical formalism does not work for the most general Horndeski theories: some terms in their field equations explicitly break the thermodynamical analogy. Strikingly, these terms are precisely those that violate the equality between the propagation speeds of gravitational and electromagnetic waves. Therefore, the crucial finding is that first-order thermodynamics indicates the direction of the physical constraints on Horndeski gravity, which paves the way for intriguing further developments. 
The analogy is spoiled for those operators which contain derivative nonminimal couplings and nonlinear contributions in the connection. This relates to the well-known but hard to tackle problem of separating matter from gravity degrees of freedom in terms of a local description.

The most general Lagrangian of Horndeski gravity reads \cite{Kobayashi:2019hrl}
\be
\mathcal{L} = \mathcal{L}_2 + \mathcal{L}_3 + \mathcal{L}_4 + 
\mathcal{L}_5 \,,
\ee
where 
\begin{align*}
\mathcal{L}_2 &= G_2 \\
\mathcal{L}_3 &= - G_3  \Box \phi\\
\mathcal{L}_4 &= G_4 \, R + G_{4 X} \left[ (\Box \phi)^2 - 
(\nabla_a \nabla_b \phi)^2 \right] \\ 
\mathcal{L}_5 &= G_5 \, G_{ab} \nabla^a \nabla^b 
\phi -  \frac{G_{5X}}{6} \Big[  (\Box \phi)^3   - 3 \, \Box \phi \, (\nabla_a \nabla_b \phi)^2 
+ 2 \, (\nabla_a \nabla_b \phi)^3 \Big]. 
\end{align*}
The $G_i (\phi, X)$ with $i=2,3,4,5$ are generic coupling functions, $G_{i\phi} \equiv \partial G_i / \partial  \phi$ and $G_{iX} \equiv \partial G_i / \partial  X$.
The viable subclass of Horndeski theories that restricts to a luminal propagation of gravitational waves is given by the choices $G_{4X}=0$ and $G_{5}=0$.

Building the thermodynamical analogy through Eckart's constitutive equations \eqref{EckartPvis}, \eqref{Eckartq} and \eqref{Eckartpi} starting from the Horndeski Lagrangian (and making the choice of neglecting bulk viscosity), one obtains
\be
\label{E-2}
\mathcal{K} \mathcal{T} = \frac{\sqrt{2 X} (G_{4 \phi} - X G_{3 
X})}{G_{4}}  \, ,
\ee
for the product between conductivity and temperature and
\be
\eta = - \frac{\sqrt{X} \, G_{4 \phi}}{\sqrt{2} \, G_{4}} 
\ee
for the shear viscosity coefficient. These expressions would of course reduce to \eqref{temperature} and \eqref{eta}, respectively, for $G_4=8\pi\phi$ and $G_3=0$.

In \cite{Giusti:2021sku} it is shown that, whenever we try to apply the thermodynamical formalism to theories beyond the viable class considered above, whose effective stress-energy tensor contains the term
\be
T_{ab}^{(\phi)} \supset \zeta(\phi, X) \, {\cal R}_{a c b d} \nabla^c \phi \nabla^d \phi \, ,
\ee
where $\zeta(\phi, X)$ is a generic function, the Riemann tensor ${\cal R}_{a c b d}$ ends up breaking the proportionality between the traceless shear tensor $\sigma_{ab}$ and the anisotropic stress tensor $\pi_{ab}^{(\phi)}$, so that Eckart's constitutive equations \eqref{EckartPvis}, \eqref{Eckartq} and \eqref{Eckartpi} no longer hold.

These results spurred some further developments, such as the considerations in \cite{Faraoni:2023hwu} that pave the way for extending the study of first-order thermodynamics of Horndeski gravity to FLRW and Bianchi universes. The imperfect fluid analogy developed for this class of theories has also been exploited with the goal of attempting to classify Horndeski theories based on the nature of the effective fluid, specifically on its requirement to be a Newtonian fluid \cite{Miranda:2022wkz}.

\section{Conclusions and outlook}
\label{concl}

The first-order thermodynamics of scalar-tensor gravity is a recent approach that characterises GR as a 
zero-temperature equilibrium state and scalar-tensor theories as non-equilibrium states, providing an interesting picture of the space of gravity theories. This 
idea is based on 
the imperfect fluid description of scalar-tensor theories and the application of Eckart's first-order thermodynamics to this effective fluid. However, it is not the recasting into an imperfect fluid form that the formalism relies on, but the fact that the constitutive relations of Eckart's thermodynamics hold and allow one to find a positive-definite expression for the temperature of modified gravity, which is nothing but a temperature relative to GR, representing the zero-temperature state.

In this work, after introducing the formalism and reviewing the analysis of several theories and some of their solutions without finding a stable equilibrium state that can challenge the central role of GR, we presented several applications and extensions of this approach. For example, an alternative picture based on the notion of chemical potential instead of temperature, that allows to treat scalar-tensor theories in the Einstein frame, an application to cosmology that confirms the physical intuition behind the formalism, and an extension to Horndeski theories that holds only for theories respecting physical constraints imposed by astrophysical observations.

There is still much more to be done to explore the potential of first-order thermodynamics for modified gravity. For example, it is very natural to consider an extension to vector-tensor theories to see if gravity with an additional vector field still has a positive, non-zero temperature. For this purpose, we are in the process of studying the simplest of these theories, Einstein-{\ae}ther \cite{Jacobson:2000xp}, and plan to extend to Generalized Proca theories \cite{Heisenberg:2014rta}, in an attempt to widen of our map of the landscape of gravity theories. 

As a cosmological application, it would be interesting (albeit non-trivial) to consider what happens in anisotropic cosmological spacetimes, such as Bianchi spacetimes, exploring the interesting connection between the presence of viscosity and cosmological anisotropies \cite{jerome}.

The long-term goal, however, remains that of going beyond Eckart's first-order thermodynamics to overcome its limitations, for example with the second-order formalisms  by Israel and Stewart \cite{Israel1979341} and by M{\"u}ller and Ruggeri \cite{MullerRuggeri93}.

\setcounter{equation}{0}

\backmatter

\bmhead{Acknowledgments}
A.G.~is supported by the European Union's Horizon 2020 research and 
innovation programme under the Marie Sk\l{}odowska-Curie Actions (grant 
agreement No.~895648). The work of A.G. and S.G. has also been carried out in the framework of the activities of the Italian National Group of 
Mathematical Physics [Gruppo Nazionale per la Fisica Matematica (GNFM), 
Istituto Nazionale di Alta Matematica (INdAM)]. {\em The results contained in the present paper have been partially presented at WASCOM 2021.}

\section*{Declarations}
\subsection*{Competing interests}
The authors declare no competing interests.

\bibliography{sn-bibliography}


\begin{thebibliography}{48}
\ifx \bisbn   \undefined \def \bisbn  #1{ISBN #1}\fi
\ifx \binits  \undefined \def \binits#1{#1}\fi
\ifx \bauthor  \undefined \def \bauthor#1{#1}\fi
\ifx \batitle  \undefined \def \batitle#1{#1}\fi
\ifx \bjtitle  \undefined \def \bjtitle#1{#1}\fi
\ifx \bvolume  \undefined \def \bvolume#1{\textbf{#1}}\fi
\ifx \byear  \undefined \def \byear#1{#1}\fi
\ifx \bissue  \undefined \def \bissue#1{#1}\fi
\ifx \bfpage  \undefined \def \bfpage#1{#1}\fi
\ifx \blpage  \undefined \def \blpage #1{#1}\fi
\ifx \burl  \undefined \def \burl#1{\textsf{#1}}\fi
\ifx \doiurl  \undefined \def \doiurl#1{\url{https://doi.org/#1}}\fi
\ifx \betal  \undefined \def \betal{\textit{et al.}}\fi
\ifx \binstitute  \undefined \def \binstitute#1{#1}\fi
\ifx \binstitutionaled  \undefined \def \binstitutionaled#1{#1}\fi
\ifx \bctitle  \undefined \def \bctitle#1{#1}\fi
\ifx \beditor  \undefined \def \beditor#1{#1}\fi
\ifx \bpublisher  \undefined \def \bpublisher#1{#1}\fi
\ifx \bbtitle  \undefined \def \bbtitle#1{#1}\fi
\ifx \bedition  \undefined \def \bedition#1{#1}\fi
\ifx \bseriesno  \undefined \def \bseriesno#1{#1}\fi
\ifx \blocation  \undefined \def \blocation#1{#1}\fi
\ifx \bsertitle  \undefined \def \bsertitle#1{#1}\fi
\ifx \bsnm \undefined \def \bsnm#1{#1}\fi
\ifx \bsuffix \undefined \def \bsuffix#1{#1}\fi
\ifx \bparticle \undefined \def \bparticle#1{#1}\fi
\ifx \barticle \undefined \def \barticle#1{#1}\fi
\bibcommenthead
\ifx \bconfdate \undefined \def \bconfdate #1{#1}\fi
\ifx \botherref \undefined \def \botherref #1{#1}\fi
\ifx \url \undefined \def \url#1{\textsf{#1}}\fi
\ifx \bchapter \undefined \def \bchapter#1{#1}\fi
\ifx \bbook \undefined \def \bbook#1{#1}\fi
\ifx \bcomment \undefined \def \bcomment#1{#1}\fi
\ifx \oauthor \undefined \def \oauthor#1{#1}\fi
\ifx \citeauthoryear \undefined \def \citeauthoryear#1{#1}\fi
\ifx \endbibitem  \undefined \def \endbibitem {}\fi
\ifx \bconflocation  \undefined \def \bconflocation#1{#1}\fi
\ifx \arxivurl  \undefined \def \arxivurl#1{\textsf{#1}}\fi
\csname PreBibitemsHook\endcsname

\bibitem{Will:2014kxa}
\begin{barticle}
\bauthor{\bsnm{Will}, \binits{C.M.}}:
\batitle{{The Confrontation between General Relativity and Experiment}}.
\bjtitle{Living Rev. Rel.}
\bvolume{17},
\bfpage{4}
(\byear{2014})
{\href{https://arxiv.org/abs/1403.7377}{{arXiv:1403.7377}}}
{[gr-qc]}.
\doiurl{10.12942/lrr-2014-4}
\end{barticle}
\endbibitem

\bibitem{Faraoni:2010pgm}
\begin{bbook}
\bauthor{\bsnm{Faraoni}, \binits{V.}},
\bauthor{\bsnm{Capozziello}, \binits{S.}}:
\bbtitle{{Beyond Einstein Gravity: A Survey of Gravitational Theories for
  Cosmology and Astrophysics}}.
\bpublisher{Springer},
\blocation{Dordrecht}
(\byear{2011}).
\doiurl{10.1007/978-94-007-0165-6}
\end{bbook}
\endbibitem

\bibitem{Capozziello:2011et}
\begin{barticle}
\bauthor{\bsnm{Capozziello}, \binits{S.}},
\bauthor{\bsnm{De~Laurentis}, \binits{M.}}:
\batitle{{Extended Theories of Gravity}}.
\bjtitle{Phys. Rept.}
\bvolume{509},
\bfpage{167}--\blpage{321}
(\byear{2011})
{\href{https://arxiv.org/abs/1108.6266}{{arXiv:1108.6266}}}
{[gr-qc]}.
\doiurl{10.1016/j.physrep.2011.09.003}
\end{barticle}
\endbibitem

\bibitem{Brans:1961sx}
\begin{barticle}
\bauthor{\bsnm{Brans}, \binits{C.}},
\bauthor{\bsnm{Dicke}, \binits{R.H.}}:
\batitle{{Mach's principle and a relativistic theory of gravitation}}.
\bjtitle{Phys. Rev.}
\bvolume{124},
\bfpage{925}--\blpage{935}
(\byear{1961}).
\doiurl{10.1103/PhysRev.124.925}
\end{barticle}
\endbibitem

\bibitem{Bergmann:1968ve}
\begin{barticle}
\bauthor{\bsnm{Bergmann}, \binits{P.G.}}:
\batitle{{Comments on the scalar tensor theory}}.
\bjtitle{Int. J. Theor. Phys.}
\bvolume{1},
\bfpage{25}--\blpage{36}
(\byear{1968}).
\doiurl{10.1007/BF00668828}
\end{barticle}
\endbibitem

\bibitem{Nordtvedt:1968qs}
\begin{barticle}
\bauthor{\bsnm{Nordtvedt}, \binits{K.}}:
\batitle{{Equivalence Principle for Massive Bodies. 2. Theory}}.
\bjtitle{Phys. Rev.}
\bvolume{169},
\bfpage{1017}--\blpage{1025}
(\byear{1968}).
\doiurl{10.1103/PhysRev.169.1017}
\end{barticle}
\endbibitem

\bibitem{Wagoner:1970vr}
\begin{barticle}
\bauthor{\bsnm{Wagoner}, \binits{R.V.}}:
\batitle{{Scalar tensor theory and gravitational waves}}.
\bjtitle{Phys. Rev. D}
\bvolume{1},
\bfpage{3209}--\blpage{3216}
(\byear{1970}).
\doiurl{10.1103/PhysRevD.1.3209}
\end{barticle}
\endbibitem

\bibitem{Faraoni:2021lfc}
\begin{barticle}
\bauthor{\bsnm{Faraoni}, \binits{V.}},
\bauthor{\bsnm{Giusti}, \binits{A.}}:
\batitle{{Thermodynamics of scalar-tensor gravity}}.
\bjtitle{Phys. Rev. D}
\bvolume{103}(\bissue{12}),
\bfpage{121501}
(\byear{2021})
{\href{https://arxiv.org/abs/2103.05389}{{arXiv:2103.05389}}}
{[gr-qc]}.
\doiurl{10.1103/PhysRevD.103.L121501}
\end{barticle}
\endbibitem

\bibitem{Giusti:2021sku}
\begin{barticle}
\bauthor{\bsnm{Giusti}, \binits{A.}},
\bauthor{\bsnm{Zentarra}, \binits{S.}},
\bauthor{\bsnm{Heisenberg}, \binits{L.}},
\bauthor{\bsnm{Faraoni}, \binits{V.}}:
\batitle{{First-order thermodynamics of Horndeski gravity}}.
\bjtitle{Phys. Rev. D}
\bvolume{105}(\bissue{12}),
\bfpage{124011}
(\byear{2022})
{\href{https://arxiv.org/abs/2108.10706}{{arXiv:2108.10706}}}
{[gr-qc]}.
\doiurl{10.1103/PhysRevD.105.124011}
\end{barticle}
\endbibitem

\bibitem{Faraoni:2021jri}
\begin{barticle}
\bauthor{\bsnm{Faraoni}, \binits{V.}},
\bauthor{\bsnm{Giusti}, \binits{A.}},
\bauthor{\bsnm{Mentrelli}, \binits{A.}}:
\batitle{{New approach to the thermodynamics of scalar-tensor gravity}}.
\bjtitle{Phys. Rev. D}
\bvolume{104}(\bissue{12}),
\bfpage{124031}
(\byear{2021})
{\href{https://arxiv.org/abs/2110.02368}{{arXiv:2110.02368}}}
{[gr-qc]}.
\doiurl{10.1103/PhysRevD.104.124031}
\end{barticle}
\endbibitem

\bibitem{Giardino:2022sdv}
\begin{barticle}
\bauthor{\bsnm{Giardino}, \binits{S.}},
\bauthor{\bsnm{Faraoni}, \binits{V.}},
\bauthor{\bsnm{Giusti}, \binits{A.}}:
\batitle{{First-order thermodynamics of scalar-tensor cosmology}}.
\bjtitle{JCAP}
\bvolume{04}(\bissue{04}),
\bfpage{053}
(\byear{2022})
{\href{https://arxiv.org/abs/2202.07393}{{arXiv:2202.07393}}}
{[gr-qc]}.
\doiurl{10.1088/1475-7516/2022/04/053}
\end{barticle}
\endbibitem

\bibitem{Faraoni:2022doe}
\begin{barticle}
\bauthor{\bsnm{Faraoni}, \binits{V.}},
\bauthor{\bsnm{Giusti}, \binits{A.}},
\bauthor{\bsnm{Jose}, \binits{S.}},
\bauthor{\bsnm{Giardino}, \binits{S.}}:
\batitle{{Peculiar thermal states in the first-order thermodynamics of
  gravity}}.
\bjtitle{Phys. Rev. D}
\bvolume{106}(\bissue{2}),
\bfpage{024049}
(\byear{2022})
{\href{https://arxiv.org/abs/2206.02046}{{arXiv:2206.02046}}}
{[gr-qc]}.
\doiurl{10.1103/PhysRevD.106.024049}
\end{barticle}
\endbibitem

\bibitem{Faraoni:2022gry}
\begin{botherref}
\oauthor{\bsnm{Faraoni}, \binits{V.}},
\oauthor{\bsnm{Giardino}, \binits{S.}},
\oauthor{\bsnm{Giusti}, \binits{A.}},
\oauthor{\bsnm{Vanderwee}, \binits{R.}}:
{Scalar field as a perfect fluid: thermodynamics of minimally coupled scalars
  and Einstein frame scalar-tensor gravity}
(2022)
{\href{https://arxiv.org/abs/2208.04051}{{arXiv:2208.04051}}}
{[gr-qc]}
\end{botherref}
\endbibitem

\bibitem{Giusti:2022tgq}
\begin{barticle}
\bauthor{\bsnm{Giusti}, \binits{A.}},
\bauthor{\bsnm{Giardino}, \binits{S.}},
\bauthor{\bsnm{Faraoni}, \binits{V.}}:
\batitle{{Past-directed scalar field gradients and scalar-tensor
  thermodynamics}}.
\bjtitle{Gen. Rel. Grav.}
\bvolume{55}(\bissue{3}),
\bfpage{47}
(\byear{2023})
{\href{https://arxiv.org/abs/2210.15348}{{arXiv:2210.15348}}}
{[gr-qc]}.
\doiurl{10.1007/s10714-023-03095-7}
\end{barticle}
\endbibitem

\bibitem{Giardino:2023qlu}
\begin{botherref}
\oauthor{\bsnm{Giardino}, \binits{S.}},
\oauthor{\bsnm{Giusti}, \binits{A.}},
\oauthor{\bsnm{Faraoni}, \binits{V.}}:
{Thermal stability of stealth and de Sitter spacetimes in scalar-tensor
  gravity}
(2023)
{\href{https://arxiv.org/abs/2302.08550}{{arXiv:2302.08550}}}
{[gr-qc]}
\end{botherref}
\endbibitem

\bibitem{Faraoni:2018qdr}
\begin{barticle}
\bauthor{\bsnm{Faraoni}, \binits{V.}},
\bauthor{\bsnm{Cot\'e}, \binits{J.}}:
\batitle{{Imperfect fluid description of modified gravities}}.
\bjtitle{Phys. Rev. D}
\bvolume{98}(\bissue{8}),
\bfpage{084019}
(\byear{2018})
{\href{https://arxiv.org/abs/1808.02427}{{arXiv:1808.02427}}}
{[gr-qc]}.
\doiurl{10.1103/PhysRevD.98.084019}
\end{barticle}
\endbibitem

\bibitem{Pimentel:1989bm}
\begin{barticle}
\bauthor{\bsnm{Pimentel}, \binits{L.O.}}:
\batitle{{Energy Momentum Tensor in the General Scalar - Tensor Theory}}.
\bjtitle{Class. Quant. Grav.}
\bvolume{6},
\bfpage{263}--\blpage{265}
(\byear{1989}).
\doiurl{10.1088/0264-9381/6/12/005}
\end{barticle}
\endbibitem

\bibitem{Eckart:1940te}
\begin{barticle}
\bauthor{\bsnm{Eckart}, \binits{C.}}:
\batitle{{The Thermodynamics of irreversible processes. 3.. Relativistic theory
  of the simple fluid}}.
\bjtitle{Phys. Rev.}
\bvolume{58},
\bfpage{919}--\blpage{924}
(\byear{1940}).
\doiurl{10.1103/PhysRev.58.919}
\end{barticle}
\endbibitem

\bibitem{Pujolas:2011he}
\begin{barticle}
\bauthor{\bsnm{Pujolas}, \binits{O.}},
\bauthor{\bsnm{Sawicki}, \binits{I.}},
\bauthor{\bsnm{Vikman}, \binits{A.}}:
\batitle{{The Imperfect Fluid behind Kinetic Gravity Braiding}}.
\bjtitle{JHEP}
\bvolume{11},
\bfpage{156}
(\byear{2011})
{\href{https://arxiv.org/abs/1103.5360}{{arXiv:1103.5360}}}
{[hep-th]}.
\doiurl{10.1007/JHEP11(2011)156}
\end{barticle}
\endbibitem

\bibitem{Miranda:2022wkz}
\begin{botherref}
\oauthor{\bsnm{Miranda}, \binits{M.}},
\oauthor{\bsnm{Vernieri}, \binits{D.}},
\oauthor{\bsnm{Capozziello}, \binits{S.}},
\oauthor{\bsnm{Faraoni}, \binits{V.}}:
{Fluid nature constrains Horndeski gravity}
(2022)
{\href{https://arxiv.org/abs/2209.02727}{{arXiv:2209.02727}}}
{[gr-qc]}
\end{botherref}
\endbibitem

\bibitem{Jacobson:1995ab}
\begin{barticle}
\bauthor{\bsnm{Jacobson}, \binits{T.}}:
\batitle{{Thermodynamics of space-time: The Einstein equation of state}}.
\bjtitle{Phys. Rev. Lett.}
\bvolume{75},
\bfpage{1260}--\blpage{1263}
(\byear{1995})
{\href{https://arxiv.org/abs/gr-qc/9504004}{{arXiv:gr-qc/9504004}}}.
\doiurl{10.1103/PhysRevLett.75.1260}
\end{barticle}
\endbibitem

\bibitem{Eling:2006aw}
\begin{barticle}
\bauthor{\bsnm{Eling}, \binits{C.}},
\bauthor{\bsnm{Guedens}, \binits{R.}},
\bauthor{\bsnm{Jacobson}, \binits{T.}}:
\batitle{{Non-equilibrium thermodynamics of spacetime}}.
\bjtitle{Phys. Rev. Lett.}
\bvolume{96},
\bfpage{121301}
(\byear{2006})
{\href{https://arxiv.org/abs/gr-qc/0602001}{{arXiv:gr-qc/0602001}}}.
\doiurl{10.1103/PhysRevLett.96.121301}
\end{barticle}
\endbibitem

\bibitem{Sotiriou:2008rp}
\begin{barticle}
\bauthor{\bsnm{Sotiriou}, \binits{T.P.}},
\bauthor{\bsnm{Faraoni}, \binits{V.}}:
\batitle{{f(R) Theories Of Gravity}}.
\bjtitle{Rev. Mod. Phys.}
\bvolume{82},
\bfpage{451}--\blpage{497}
(\byear{2010})
{\href{https://arxiv.org/abs/0805.1726}{{arXiv:0805.1726}}}
{[gr-qc]}.
\doiurl{10.1103/RevModPhys.82.451}
\end{barticle}
\endbibitem

\bibitem{Waldbook}
\begin{bbook}
\bauthor{\bsnm{Wald}, \binits{R.M.}}:
\bbtitle{{General Relativity}}.
\bpublisher{Chicago Univ. Press},
\blocation{Chicago, USA}
(\byear{1984}).
\doiurl{10.7208/chicago/9780226870373.001.0001}
\end{bbook}
\endbibitem

\bibitem{Faraoni:2023hwu}
\begin{botherref}
\oauthor{\bsnm{Faraoni}, \binits{V.}},
\oauthor{\bsnm{Houle}, \binits{J.}}:
{More on the first-order thermodynamics of scalar-tensor and Horndeski gravity}
(2023)
{\href{https://arxiv.org/abs/2302.01442}{{arXiv:2302.01442}}}
{[gr-qc]}
\end{botherref}
\endbibitem

\bibitem{Maartens:1996vi}
\begin{bchapter}
\bauthor{\bsnm{Maartens}, \binits{R.}}:
\bctitle{{Causal thermodynamics in relativity}}.
(\byear{1996})
\end{bchapter}
\endbibitem

\bibitem{Afshordi:2006ad}
\begin{barticle}
\bauthor{\bsnm{Afshordi}, \binits{N.}},
\bauthor{\bsnm{Chung}, \binits{D.J.H.}},
\bauthor{\bsnm{Geshnizjani}, \binits{G.}}:
\batitle{{Cuscuton: A Causal Field Theory with an Infinite Speed of Sound}}.
\bjtitle{Phys. Rev. D}
\bvolume{75},
\bfpage{083513}
(\byear{2007})
{\href{https://arxiv.org/abs/hep-th/0609150}{{arXiv:hep-th/0609150}}}.
\doiurl{10.1103/PhysRevD.75.083513}
\end{barticle}
\endbibitem

\bibitem{nordstroem}
\begin{barticle}
\bauthor{\bsnm{Nordström}, \binits{G.}}:
\batitle{{Zur Theorie der Gravitation vom Standpunkt des
  Relativit\"atsprinzips}}.
\bjtitle{Annalen der Physik}
\bvolume{347}(\bissue{13}),
\bfpage{533}--\blpage{554}
(\byear{1913}).
\doiurl{10.1002/andp.19133471303}
\end{barticle}
\endbibitem

\bibitem{Afshordi:2009tt}
\begin{barticle}
\bauthor{\bsnm{Afshordi}, \binits{N.}}:
\batitle{{Cuscuton and low energy limit of Horava-Lifshitz gravity}}.
\bjtitle{Phys. Rev. D}
\bvolume{80},
\bfpage{081502}
(\byear{2009})
{\href{https://arxiv.org/abs/0907.5201}{{arXiv:0907.5201}}}
{[hep-th]}.
\doiurl{10.1103/PhysRevD.80.081502}
\end{barticle}
\endbibitem

\bibitem{Afshordi:2007yx}
\begin{barticle}
\bauthor{\bsnm{Afshordi}, \binits{N.}},
\bauthor{\bsnm{Chung}, \binits{D.J.H.}},
\bauthor{\bsnm{Doran}, \binits{M.}},
\bauthor{\bsnm{Geshnizjani}, \binits{G.}}:
\batitle{{Cuscuton Cosmology: Dark Energy meets Modified Gravity}}.
\bjtitle{Phys. Rev. D}
\bvolume{75},
\bfpage{123509}
(\byear{2007})
{\href{https://arxiv.org/abs/astro-ph/0702002}{{arXiv:astro-ph/0702002}}}.
\doiurl{10.1103/PhysRevD.75.123509}
\end{barticle}
\endbibitem

\bibitem{Quintin:2019orx}
\begin{barticle}
\bauthor{\bsnm{Quintin}, \binits{J.}},
\bauthor{\bsnm{Yoshida}, \binits{D.}}:
\batitle{{Cuscuton gravity as a classically stable limiting curvature theory}}.
\bjtitle{JCAP}
\bvolume{02},
\bfpage{016}
(\byear{2020})
{\href{https://arxiv.org/abs/1911.06040}{{arXiv:1911.06040}}}
{[gr-qc]}.
\doiurl{10.1088/1475-7516/2020/02/016}
\end{barticle}
\endbibitem

\bibitem{Faraoni:2022jyd}
\begin{barticle}
\bauthor{\bsnm{Faraoni}, \binits{V.}},
\bauthor{\bsnm{Fran\c{c}onnet}, \binits{T.B.}}:
\batitle{{Stealth metastable state of scalar-tensor thermodynamics}}.
\bjtitle{Phys. Rev. D}
\bvolume{105}(\bissue{10}),
\bfpage{104006}
(\byear{2022})
{\href{https://arxiv.org/abs/2203.14934}{{arXiv:2203.14934}}}
{[gr-qc]}.
\doiurl{10.1103/PhysRevD.105.104006}
\end{barticle}
\endbibitem

\bibitem{Bardeen:1980kt}
\begin{barticle}
\bauthor{\bsnm{Bardeen}, \binits{J.M.}}:
\batitle{{Gauge Invariant Cosmological Perturbations}}.
\bjtitle{Phys. Rev. D}
\bvolume{22},
\bfpage{1882}--\blpage{1905}
(\byear{1980}).
\doiurl{10.1103/PhysRevD.22.1882}
\end{barticle}
\endbibitem

\bibitem{Ellis:1989jt}
\begin{barticle}
\bauthor{\bsnm{Ellis}, \binits{G.F.R.}},
\bauthor{\bsnm{Bruni}, \binits{M.}}:
\batitle{{Covariant and Gauge Invariant Approach to Cosmological Density
  Fluctuations}}.
\bjtitle{Phys. Rev. D}
\bvolume{40},
\bfpage{1804}--\blpage{1818}
(\byear{1989}).
\doiurl{10.1103/PhysRevD.40.1804}
\end{barticle}
\endbibitem

\bibitem{Ellis:1990gi}
\begin{barticle}
\bauthor{\bsnm{Ellis}, \binits{G.F.R.}},
\bauthor{\bsnm{Bruni}, \binits{M.}},
\bauthor{\bsnm{Hwang}, \binits{J.}}:
\batitle{{Density Gradient - Vorticity Relation in Perfect Fluid
  {Robertson-Walker} Perturbations}}.
\bjtitle{Phys. Rev. D}
\bvolume{42},
\bfpage{1035}--\blpage{1046}
(\byear{1990}).
\doiurl{10.1103/PhysRevD.42.1035}
\end{barticle}
\endbibitem

\bibitem{Hwang:1990am}
\begin{barticle}
\bauthor{\bsnm{Hwang}, \binits{J.-c.}},
\bauthor{\bsnm{Vishniac}, \binits{E.T.}}:
\batitle{{Analyzing cosmological perturbations using the covariant approach}}.
\bjtitle{Astrophys. J.}
\bvolume{353},
\bfpage{1}--\blpage{20}
(\byear{1990}).
\doiurl{10.1086/168583}
\end{barticle}
\endbibitem

\bibitem{Miranda:2022uyk}
\begin{barticle}
\bauthor{\bsnm{Miranda}, \binits{M.}},
\bauthor{\bsnm{Graham}, \binits{P.-A.}},
\bauthor{\bsnm{Faraoni}, \binits{V.}}:
\batitle{{Effective fluid mixture of tensor-multi-scalar gravity}}.
\bjtitle{Eur. Phys. J. Plus}
\bvolume{138}(\bissue{5}),
\bfpage{387}
(\byear{2023})
{\href{https://arxiv.org/abs/2211.03958}{{arXiv:2211.03958}}}
{[gr-qc]}.
\doiurl{10.1140/epjp/s13360-023-03984-5}
\end{barticle}
\endbibitem

\bibitem{mybook1}
\begin{bbook}
\bauthor{\bsnm{Faraoni}, \binits{V.}}:
\bbtitle{{Cosmology in Scalar-Tensor Gravity}}.
\bpublisher{Kluwer Academic},
\blocation{Dordrecht, The Netherlands}
(\byear{2004}).
\doiurl{10.1007/978-1-4020-1989-0}
\end{bbook}
\endbibitem

\bibitem{Nariai}
\begin{barticle}
\bauthor{\bsnm{Nariai}, \binits{H.}}:
\batitle{{On the Green's Function in an Expanding Universe and Its Role in the
  Problem of Mach's Principle}}.
\bjtitle{Progress of Theoretical Physics}
\bvolume{40}(\bissue{1}),
\bfpage{49}--\blpage{59}
(\byear{1968}).
\doiurl{10.1143/PTP.40.49}
\end{barticle}
\endbibitem

\bibitem{Faraoni:2022fxo}
\begin{barticle}
\bauthor{\bsnm{Faraoni}, \binits{V.}},
\bauthor{\bsnm{Graham}, \binits{P.-A.}},
\bauthor{\bsnm{Leblanc}, \binits{A.}}:
\batitle{{Critical solutions of nonminimally coupled scalar field theory and
  first-order thermodynamics of gravity}}.
\bjtitle{Phys. Rev. D}
\bvolume{106}(\bissue{8}),
\bfpage{084008}
(\byear{2022})
{\href{https://arxiv.org/abs/2207.03841}{{arXiv:2207.03841}}}
{[gr-qc]}.
\doiurl{10.1103/PhysRevD.106.084008}
\end{barticle}
\endbibitem

\bibitem{Faraoni:2003jh}
\begin{barticle}
\bauthor{\bsnm{Faraoni}, \binits{V.}}:
\batitle{{Possible end of the universe in a finite future from dark energy with
  w-1}}.
\bjtitle{Phys. Rev. D}
\bvolume{68},
\bfpage{063508}
(\byear{2003})
{\href{https://arxiv.org/abs/gr-qc/0307086}{{arXiv:gr-qc/0307086}}}.
\doiurl{10.1103/PhysRevD.68.063508}
\end{barticle}
\endbibitem

\bibitem{Kobayashi:2019hrl}
\begin{barticle}
\bauthor{\bsnm{Kobayashi}, \binits{T.}}:
\batitle{{Horndeski theory and beyond: a review}}.
\bjtitle{Rept. Prog. Phys.}
\bvolume{82}(\bissue{8}),
\bfpage{086901}
(\byear{2019})
{\href{https://arxiv.org/abs/1901.07183}{{arXiv:1901.07183}}}
{[gr-qc]}.
\doiurl{10.1088/1361-6633/ab2429}
\end{barticle}
\endbibitem

\bibitem{Ezquiaga:2018btd}
\begin{barticle}
\bauthor{\bsnm{Ezquiaga}, \binits{J.M.}},
\bauthor{\bsnm{Zumalac\'arregui}, \binits{M.}}:
\batitle{{Dark Energy in light of Multi-Messenger Gravitational-Wave
  astronomy}}.
\bjtitle{Front. Astron. Space Sci.}
\bvolume{5},
\bfpage{44}
(\byear{2018})
{\href{https://arxiv.org/abs/1807.09241}{{arXiv:1807.09241}}}
{[astro-ph.CO]}.
\doiurl{10.3389/fspas.2018.00044}
\end{barticle}
\endbibitem

\bibitem{Jacobson:2000xp}
\begin{barticle}
\bauthor{\bsnm{Jacobson}, \binits{T.}},
\bauthor{\bsnm{Mattingly}, \binits{D.}}:
\batitle{{Gravity with a dynamical preferred frame}}.
\bjtitle{Phys. Rev. D}
\bvolume{64},
\bfpage{024028}
(\byear{2001})
{\href{https://arxiv.org/abs/gr-qc/0007031}{{arXiv:gr-qc/0007031}}}.
\doiurl{10.1103/PhysRevD.64.024028}
\end{barticle}
\endbibitem

\bibitem{Heisenberg:2014rta}
\begin{barticle}
\bauthor{\bsnm{Heisenberg}, \binits{L.}}:
\batitle{{Generalization of the Proca Action}}.
\bjtitle{JCAP}
\bvolume{05},
\bfpage{015}
(\byear{2014})
{\href{https://arxiv.org/abs/1402.7026}{{arXiv:1402.7026}}}
{[hep-th]}.
\doiurl{10.1088/1475-7516/2014/05/015}
\end{barticle}
\endbibitem

\bibitem{jerome}
\begin{barticle}
\bauthor{\bsnm{Ganguly}, \binits{C.}},
\bauthor{\bsnm{Quintin}, \binits{J.}}:
\batitle{{Microphysical manifestations of viscosity and consequences for
  anisotropies in the very early universe}}.
\bjtitle{Phys. Rev. D}
\bvolume{105}(\bissue{2}),
\bfpage{023532}
(\byear{2022})
{\href{https://arxiv.org/abs/2109.11701}{{arXiv:2109.11701}}}
{[gr-qc]}.
\doiurl{10.1103/PhysRevD.105.023532}
\end{barticle}
\endbibitem

\bibitem{Israel1979341}
\begin{barticle}
\bauthor{\bsnm{Israel}, \binits{W.}},
\bauthor{\bsnm{Stewart}, \binits{J.M.}}:
\batitle{Transient relativistic thermodynamics and kinetic theory}.
\bjtitle{Annals of Physics}
\bvolume{118}(\bissue{2}),
\bfpage{341}--\blpage{372}
(\byear{1979}).
\doiurl{10.1016/0003-4916(79)90130-1}
\end{barticle}
\endbibitem

\bibitem{MullerRuggeri93}
\begin{bbook}
\bauthor{\bsnm{M{\"u}ller}, \binits{I.}},
\bauthor{\bsnm{Ruggeri}, \binits{T.}}:
\bbtitle{{Extended Thermodynamics}}.
\bpublisher{Springer},
\blocation{New York, USA}
(\byear{1993}).
\doiurl{10.1007/978-1-4684-0447-0}
\end{bbook}
\endbibitem

\end{thebibliography}


\end{document}